\documentclass[twocolumn,showpacs,preprintnumbers,amsmath,amssymb]{revtex4}
\usepackage{epsfig,epsf,graphics,color,rotate}
\usepackage{dcolumn}
\usepackage{bm}
\usepackage{amsmath}
\usepackage{amsfonts}
\usepackage{fancyhdr}
\usepackage{psfrag}
\newcommand\Tr{\mbox{Tr}}

\begin{document}



\title{Bounds on the attractor dimension for low$-Rm$ wall-bound MHD turbulence}

\author{Alban Poth\'erat$^1$ and Thierry Alboussi\`ere$^2$}
\email{alban.potherat@tu-ilmenau.de}
\affiliation{$^1$Ilmenau Technical University, Kirchoffstr. 1 98693 Ilmenau, Germany\\
$^2$Laboratoire de G\'eophysique Interne et Tectonophysique, CNRS, Observatoire de Grenoble, Universit\'e Joseph Fourier \\ Maison des G\'eosciences, BP 53, 38041 Grenoble Cedex 9, France}
\email{talbouss@obs.ujf-grenoble.fr}
\date{13 October, 2006}

\begin{abstract}
Steady Low $R_m$ MHD turbulence is investigated here through estimates of upper 
bounds for attractor dimension. A flow between two parallel walls with an imposed perpendicular magnetic field is considered. The flow is defined by its maximum velocity and the intensity of the magnetic field. Given the corresponding Reynolds and Hartmann numbers, one can rigorously derive an upper bound for the dimension of the attractor and find out which modes must be chosen to achieve this bound. The properties of these modes yield quantities which we compare 
to known heuristics estimates for  the size of 
the smallest turbulent vortices and the degree of anisotropy of the turbulence. 
Our upper bound derivation is based on 
known bounds of the non-linear inertial term, while low $R_m$ Lorentz forces -- being linear -- can be relatively easily dealt with.

The simple configuration considered in this paper allows us to identify some 
boundaries separating different sets of modes in the space of non-dimensional 
parameters, which are reminiscent of three important previously identified 
transitions observed in the real flow. The first boundary separate  
classical hydrodynamic sets of modes from  MHD sets where
anisotropy takes the form of a ``Joule cone''. In the  second, one can define 
the boundary separating 3D MHD sets from quasi-2D MHD sets, when all 
``Sommerfeld modes'' disappear and only ``Squire modes'' are left. The third separates sets where all the modes exhibit the same boundary layer thickness or so, and sets where
 many different ``boundary layer modes'' co-exists in the set. The 
non-dimensional relations defining these boundaries are 
then compared to the heuristics known for the transition between isotropic and 
anisotropic MHD turbulence, 3D and quasi-2D MHD turbulence and that between a 
turbulent and a laminar Hartmann layer.

In addition to this 3D approach, we also determine upper bounds for the 
dimension of forced turbulent flows modelled using a 2D MHD equation, which 
should become physically relevant in the quasi-2D MHD regime. The advantage of 
this 2D approach is that, while upper bounds are quite loose in three 
dimensions, optimal upper bounds exist for the 2D nonlinear term. This allows us to derive realistic attractor dimensions for quasi-2D MHD flows.  

\end{abstract}
\maketitle
\section{Introduction}
\label{intro}

%
%

Magnetohydrodynamic turbulence at low magnetic Reynolds number is relevant to 
many laboratory-scale experiments \cite{alemany79,hua74,sommeria86,psm05} as well as to many industrial 
processes \cite{buhler01,wolf94}. This is a consequence of the small values of liquid metal Prandtl numbers $Pm$, around $10^{-6}$. With  magnetic fields in the range 0.1 to 1 Tesla, lengthscales of 0.1 to 1 m and flow velocities of 0.01 
to 1 m/s, it is likely that Lorentz forces will have a great impact on liquid metal flow turbulence and unlikely that the flow will affect the imposed magnetic field, as magnetic diffusivity exceeds kinematic viscosity by a factor $Pm^{-1}=10^6$.  Unexpectedly, magnetohydrodynamic turbulence can also be relevant to large magnetic Reynolds numbers when the magnetic Prandtl number is small. This is the case for the geodynamo, for which the magnetic Prandtl number is about $10^{-6}$. In this case, going from large eddies to smaller ones, the magnetic turbulent cascade stops long before the hydrodynamic cascade. The small scales of large-$Rm$ turbulence can thus be considered as an example of low-$Rm$ turbulence. \\
The question of how turbulence looks can be addressed theoretically in a 
number of ways, most of them based on physical assumptions or turbulence models
as recently in \cite{ueno06}.
 It is, however, possible to make no assumption and although this leads to 
 limited information, this is reliable information on which further studies 
 can be based, for instance the derivation of a turbulence model. In this paper,
 we have estimated upper bounds for the attractor dimension of low-Rm MHD turbulent flows, a quantity that can be rigorously linked to the size of the 
dissipative scales \cite{constantin85_jfm} and therefore provides some crucial 
information on the flow. Our work is based on previous non-MHD studies of Navier-Stokes attractors in three and two dimensions. It has been shown that the attractor of turbulent solutions is of finite dimension even though the set of possible functions is of infinite dimension. This is basically due to viscous effects setting a maximum curvature to possible solutions. In three dimensions, the fundamental questions of the smoothness of solutions of Navier-Stokes equations and of the existence of a compact attractor remain open. We assume this is the case and this constitutes the only assumption. We do not need to make such assumptions for the two-dimensional case as smoothness has been proved. In addition, previous upper bounds of attractor dimensions for two-dimensional flows coincide 
very nearly with heuristic estimates based on the cascade of enstrophy \cite{kraich67,constantin85_jfm}. This is not the case for three dimensional rigorous upper bounds as they are found to be much larger than heuristic estimates based on the cascade of energy \cite{k41,constantin85_jfm}. \\

Our main goal is basically to extend the non-MHD studies to the case of low-$Rm$ MHD. As Lorentz forces are linear forces, this does not include additional theoretical difficulties. However, as these forces are strongly dissipative, the final upper bounds for attractor dimensions are significantly affected when imposing a magnetic field. Together with the upper bounds, we put in evidence the modes that achieve the upper bound for the attractor dimension. We then determine 
 boundaries separating different sets of these modes exhibiting the
same properties as real turbulent flows. These boundaries are 
encountered when increasing the imposed magnetic field and strongly resemble 
known transitions occurring in the flow. The chosen 
configuration of study is taken as simple as possible to enable us to show 
these transitions: the electrically conducting fluid is contained between two 
parallel infinite walls, and a perpendicular magnetic field is applied. The 
flow is held steady on average by application of a forcing which is independent 
of the velocity 
field. First, it is expected that isotropic turbulence becomes anisotropic by stretching in the direction of the magnetic field \cite{moffatt67}. Secondly, the turbulent 
Hartmann boundary layers which develop along the walls become laminar and 
thirdly,  a quasi two-dimensional regime is obtained. \\

Three similar boundaries between sets of modes can be observed and determined 
formally by analysing rigorous upper bounds for attractor dimension and the corresponding set of vector-valued functions. The first one is defined by the 
absence of functions with variations along the magnetic field direction outside (Hartmann) boundary layers and with no variation in the perpendicular directions. The third boundary is found when this absence is independent of any condition regarding perpendicular
 variations. Finally the second boundary corresponds to the presence modes with  other length-scales close to the walls than the laminar Hartmann layer 
thickness.\\


The method of estimates of dimension upper bounds for attractors is presented 
in section \ref{sec:method}. Section \ref{sec:modes} is devoted to the 
determination of least dissipative modes. In section \ref{sec:dim3D} and 
\ref{sec:q2D} we determine the attractor dimension of three dimensional and 
two dimensional turbulent flows respectively and discuss the various boundaries found, and compare them to the transitions  observed as the nature of 
turbulence changes with the intensity of the magnetic field. Comments on this 
work and results are presented in the final section \ref{sec:concl}, where the 
relevance of the least dissipative modes to the real flow is discussed.  


\section{Method to determine dimension and modes of the attractor}
\label{sec:method}

From the point of view of dynamical systems, the phase space of turbulence is the infinite-dimensional set of all vector-valued functions in which Navier-Stokes solutions evolve as time goes on. If the attractor is compact, its finite dimension $d$ can be determined using the fact that any solution to the Navier-Stokes equations should eventually become arbitrarily close to the attractor. Let us then consider an infinitesimal $n$-dimensional box around a solution. As time passes, each point in the box evolves according to the Navier-Stokes equations. 
Eventually, this box will then end up within the attractor. This provides a 
criterion on the fate of the volume of the box: if $n>d$ the volume of the box will converge towards zero, in the same way as a cube has to have a vanishing volume if it is to fit in a surface of zero thickness. As soon as $n<d$ the volume of the box will not converge towards zero, {\it i.e.} it will either not converge,  converge towards a finite value, or diverge -- should ergodicity hold -- as the box spreads into the whole attractor .  

The evolution of the size of $n$-dimensional infinitesimal boxes is intimately related to the Lyapunov exponents. The $n$-th Lyapunov exponent is the maximal rate of growth of the $n$-th dimensional volume spanned by $n$ infinitesimal disturbances about a solution at large time ({\it i.e.} in the attractor). Denoting ${\cal{A}}$ the linearised Navier-Stokes equations, a disturbance $\delta {\bf u}$ about a solution ${\bf u}$ obeys the following equation~:
\begin{equation}
\frac{\partial }{\partial t} \delta {\bf u} = {\cal{A}}({\bf u}) \delta {\bf u} + {\cal{O}} (\delta {\bf u} ^2 ). 
\label{linear}
\end{equation} 
Considering $n$ orthonormal disturbances spanning a $n$-volume 
$V_n = \| \delta {\bf u}_1 \times \dots \times \delta {\bf u}_n \|$, the 
equation for a single disturbance (\ref{linear}) can be generalised and 
integrated to provide an expression for $V_n$ (see \cite{doering95}):
\begin{equation}
V_n (t) = V_n (0) \exp \left( t \left< \Tr \left[ {\cal{A}} P_n  \right] \right>  \right), 
\label{volume}
\end{equation}
where $P_n$ denotes the projection onto the $n$-dimensional subspace spanned by 
the $n$ disturbances at initial time, and the bracket $< >$ stands for long 
time-average. 
For each integer value $n$, we determine the $n$ largest eigenvalues of 
$\mathcal A$ and 
consider whether their sum is larger or smaller than zero. In the first 
case, $n$ is smaller than the attractor dimension, in the second case, $n$ is larger than the attractor dimension.

Non-linear terms can stretch solutions and are pushing eigenvalues towards positive values, while dissipative terms make solutions shrink (in phase space) and lead to negative eigenvalues. The largest eigenvalues are obtained for large scale functions while small scales vector fields produce more dissipation and thus negative eigenvalues. Hence, large scale motions are selected first to calculate the first Lyapunov exponents. Their number is limited though and eventually one has to select negative eigenvalues. Eventually, when enough negative values are added up, they balance the initial positive eigenvalues and Lyapunov exponents themselves become negative. The $n$-th Lyapunov exponent for which this happens 
roughly yields  the dimension of the turbulent attractor 
\footnote{These dimensions are numbers much larger than one, so we do not need to worry very long about their exact non-integer value between $n-1$ and $n$. For low-dimensional attractors, the Kaplan-Yorke formula can be used to determine this value precisely \cite{holmes96}.  }. 

The above cross products and projection require the existence of a scalar product in the phase space of vector valued functions (in 2D or 3D space). In this study we shall use the standard $L^2$ Hilbert structure. Given 
${\bf u}$ and ${\bf v}$ two vector fields, their scalar product is defined as:
\begin{equation}
{\bf u}.{\bf v} = \int_{\cal{V}} u_i v_i^* dV, 
\label{scalar}  
\end{equation}
where $i=1,2$ or $i=1,2,3$ for 2D or 3D turbulence. 
We are considering here low Rm turbulence under the condition of
diffusive magnetic field, i.e. the quasi-static model of MHD. If one
considers a small perturbation, $\delta \mathbf u$, its evolution is thus 
governed by the following equations:
\begin{eqnarray}
\partial_t \delta \mathbf u
&=& -{\mathbf \nabla} \delta p -\mathbf u .\nabla \delta \mathbf u - \delta \mathbf u .\nabla\mathbf  u 
\nonumber \\
+ \nu (\mathbf \nabla^{2}
+\frac{\sigma B^2}{\rho \nu} \nabla^{-2} \partial^2_{zz})\delta \mathbf u 
\label{eq:ns_var}\\
\nabla \cdot \delta \mathbf u &=& 0
\end{eqnarray}
with associated boundary conditions for the perturbation:
\begin{eqnarray}
\forall \mathbf x \in \mathbb R^3, \forall k_x,k_y \in  \mathbb Z, \nonumber \\
\mathbf v(\mathbf x)= \mathbf v(\mathbf x+ k_x  L \mathbf e_x)= \mathbf v(\mathbf x+ k_y L 
\mathbf e_y)\\
\mathbf v(z=-L)=\mathbf v(z=L)=0
\label{eq:bc_perturbation}
\end{eqnarray}
where $\sigma, \rho$ and $\nu$ are the fluid's electric conductivity, density 
and kinematic viscosity respectively. $B$ is the intensity of the magnetic 
field, which points in the $z-$direction. The perturbation in electric 
current $\delta \mathbf j$ is related to $\delta \mathbf u$ by
\begin{equation}
\delta \mathbf j=-\sigma B\nabla^{-2} \partial _z \nabla \times \delta \mathbf u
\label{eq:deltaj_omega}
\end{equation}
and must satisfy the perturbed form of
the electric current conservation $\nabla \cdot \delta \mathbf j=0$ as well as
 the condition that the walls located at $z=-L$ and $z=L$ are electrically 
 insulating:
\begin{equation}
\delta j_{n}=0
\label{eq:thinw_deltaj}
\end{equation}
where $\delta j_{n}$ is the component of $\delta \mathbf j$ which is normal to
 the wall.\\
 The trace of the evolution operator is split into non-linear and linear parts, ${\cal{A}}({\bf u}) \delta {\bf u} = {\cal{B}}({\bf u},\delta {\bf u}) + {\cal{L}}(\delta {\bf u})$. For any disturbance $\delta {\bf u}$ of norm unity, the 
contribution of the non linear term to the trace operator is expressed as:
\begin{equation}
\int _{\cal{V}} \delta {\bf u}.{\cal{B}}({\bf u},\delta {\bf u}) dV = \int _{\cal{V}} \delta {\bf u}. \left[ -{\mathbf \nabla} \delta p -\mathbf u .\nabla \delta \mathbf u - \delta \mathbf u .\nabla\mathbf  u\right] dV 
\label{nonlin}
\end{equation}
It has been shown in \cite{constantin85_ams} that this contribution can be 
bound as follows:
\begin{eqnarray}
\int _{\cal{V}} \delta {\bf u}.{\cal{B}}({\bf u},\delta {\bf u}) dV & \leq & \left| \nabla \delta {\bf u} \right| _{L^2}  \vert {\mathbf u} \vert _{L^\infty}
\nonumber \\
& \leq & - \frac{\nu }{2} \int _{\cal{V}} \left( \nabla ^2 \delta {\bf u} \right) . \delta {\bf u} \  dV + \frac{1}{2 \nu } \vert {\mathbf u} \vert _{L^\infty}^2  
\label{nonlinref}
\end{eqnarray}
The first term is equal to half the viscous dissipation term with the opposite 
sign and the second term depends on the maximum velocity which we assumed to be 
bounded. 
When considering $n$ disturbances $(\delta {\bf u}_i)$ the trace of the 
linearized operator ${\cal{A}}$ on the subspace spanned by these disturbances 
can be expressed as:
\begin{equation}
\Tr({\cal{A}}P_n) = \Tr({\cal{B}}P_n) + \Tr ({\cal{L}}P_n) 
\leq \frac{\nu }{2} \Tr( \nabla ^2 P_n )   
+ \frac{\sigma B^2}{\rho} \Tr( {\nabla }^{-2} \partial ^2_{zz} P_n  ) + \frac{n}{2 \nu } \vert {\mathbf u} \vert _{L^\infty}^2 
\end{equation}
This can be be written in dimensionless terms using $L$ for distance, $U_\infty$
the maximum velocity for velocity, $\sigma B U_\infty$ for electric current 
density and $\nu / L^3$ for traces of operators:
\begin{equation}
\Tr({\cal{A}}P_n) \leq \Tr \left(\left[ \frac{1}{2} \nabla ^2+H\!a^2 {\nabla }^{-2} \partial ^2_{zz}\right]P_n  \right) + \frac{n}{2} Re^2 
\label{eq:tra}
\end{equation}
where the Hartmann and Reynolds numbers are defined as $H\!a = \sqrt{\sigma / (\rho \nu ) }B L $ and $Re = U_\infty L / \nu $. This bound can be determined provided the trace of the operator 
$\mathcal D_{Ha}=1/2 \nabla^2+Ha \nabla^{-2} \partial_{zz}$ which is nearly 
equal to the dissipative operator (viscous and Joule dissipation), can be evaluated. This is the subject of the next section.  Before we move on to this task, 
it is important to notice that although the equations for $\delta \mathbf u$ 
have been linearized around an unspecified solution $\mathbf u$, the attractor
to which $\mathbf u$ belong is indeed that of the full non-linear 
Navier Stokes equations, and so is its his dimension $d_M$.

\section{Least dissipative modes in a flow between 2 Hartmann walls}
\label{sec:modes}

In the expression for the trace of the linearized evolution operator (\ref{eq:tra}), the last term is positive and depends only on the number of modes $n$ and not on which modes are considered. The other term is the trace of a linear operator, consisting of half the viscous effect and of the entire Lorentz force. This term is negative and depends on the modes selected. In order to find an upper bound for the attractor dimension, we want to select the least dissipative terms so as to get as many of them as possible before the trace vanishes. Finding the 
least dissipative modes boils down to an eigenvalue problem. 

\subsection{Eigenvalue problem}
\label{sec:eigen}
In this section, we solve the eigenvalue problem for the dissipation operator $\mathcal D_{Ha}$ in a closed box with $L$-periodic boundary conditions in the 
$x$ and $y$ direction and impermeable and electrically insulating walls located at $z=-L$ and $z=L$. For these boundary conditions, the Laplacian operator is invertible so that the eigenvalue problem for the $\mathcal D_{Ha}$
operator can be formulated using non-dimensional variables as:
\begin{eqnarray}
(\nabla^4-2Ha^2 \partial^2_{zz})\mathbf v&=& 2\lambda\nabla^2 \mathbf v
\label{eq:eigenprob}\\
\nabla . \mathbf v&=&0
\label{eq:continuity}
\end{eqnarray}
%
%
Here lengths are normalised by $L$, velocities by an unspecified typical 
velocity $U$, the eigenvalues $\lambda$ by $\nu/L^2$ and electric 
currents by $\sigma B U$. The choice of $U$ is not important since the problem 
is linear.
$\mathbf v$ also has to satisfy the same kinematic boundary conditions 
(\ref{eq:bc_perturbation}) as $\delta \mathbf u$. As in 
(\ref{eq:deltaj_omega}), an electric current field $\mathbf J$ is associated to $\mathbf v$ by:
\begin{equation}
\mathbf J=-\nabla^{-2}\partial_z \Omega
\label{eq:J_omega}
\end{equation}
where $\Omega=\nabla \times \mathbf v$. $\mathbf J$ must be solenoidal and
satisfy the same electric boundary conditions (\ref{eq:thinw_deltaj}) 
as $\delta \mathbf j$ at the walls located at $z=-1$ and $z=1$.\\
Since $\frac{\partial}{\partial x}$, $\frac{\partial}{\partial y}$ and
$\frac{\partial}{\partial z}$ commute with $\mathcal D_{Ha}$, each component $v_x$, $v_y$ and $v_z$ of
the solution $\mathbf v=(v_i)_{i\in\{x,y,z\}}$ of (\ref{eq:eigenprob}) is of the form:
\begin{equation}
v_i(\mathbf x)=V_i\exp(i\mathbf {k_\perp. x_\perp}+\phi_i)Z_i(z)
\label{eq:form_vi}
\end{equation}
with $\phi_i \in ]-\pi, \pi]$ and  
$\mathbf k_\perp=k_x \mathbf e_x+k_y\mathbf e_y$. 
The periodic boundary conditions in the $x$ and $y$ directions impose 
$(k_x,k_y) \in 2\pi \mathbb Z^2$. 
$Z_i(z)=\sum_{j}A_i^{(j)} \exp(K^{(j)} z)$ where $K^{(j)}$ are the complex 
roots of the dispersion equation obtained by inserting (\ref{eq:form_vi}) 
into (\ref{eq:eigenprob}): 
\begin{equation}
2\lambda=-(k_x^2+k_y^2+K^2)-2Ha^2\frac{K^2}{k_x^2+k_y^2+K^2}
\label{eq:disp}
\end{equation}
There are always  two real  and two  imaginary roots for $K$: 
$1/\delta$, $-1/\delta$, $i\kappa_z$ and $-i\kappa_z$ with:
\begin{eqnarray}
-\kappa_z^2=Ha^2+\lambda+k_\perp^2-\sqrt{(Ha^2+\lambda)^2+2k_\perp^2 Ha^2}
\label{eq:kappa_z}\\
\frac{1}{\delta^2}=Ha^2+\lambda+k_\perp^2+\sqrt{(Ha^2+\lambda)^2+2k_\perp^2 Ha^2}
\label{eq:delta}
\end{eqnarray}
Eventually, each function $Z_i(z)$ is of the general form:
\begin{equation}
Z_i(z)=A_i^1 \exp(z/\delta)+A_i^2 \exp(-z/\delta)
+A_i^3 \exp(i\kappa_z z)+A_i^4 \exp(-i\kappa_z z)
\label{eq:form_zi}
\end{equation}
Since the operator $\mathcal D_{Ha}$ has homogeneous boundary conditions, 
there is a 
discrete spectrum of possible values of $\delta$ and $\kappa_z$, which is 
determined by the boundary conditions.
We shall now express the boundary conditions for the $Z_i$ functions derived in
the previous section. The impermeability conditions at $z=-1$ and $z=1$ yield
readily:
\begin{eqnarray}
Z_i(-1)=0
\label{eq:imp1}\\
Z_i(1)=0
\label{eq:imp2}
\end{eqnarray}
which, by virtue of the continuity equation (\ref{eq:continuity}) implies:
\begin{eqnarray}
Z_z'(-1)=0
\label{eq:cont1}\\
Z_z'(1)=0
\label{eq:cont2}
\end{eqnarray}


Using (\ref{eq:J_omega}) to calculate $J_z(z=-1)$ and $J_z(z=1)$, 
 then using the dispersion equation (\ref{eq:disp}), 
the electric conditions at the walls located at $z=-1$ and $z=1$ are 
respectively written:
\begin{equation}
-k_x\left(Z_y^\prime(-1)-(k_\perp^2+2\lambda)\int Z_y(-1) \right)
+k_y\left(Z_x^\prime(-1)-(k_\perp^2+2\lambda)\int Z_x(-1)\right)=0
\label{eq:elec1}
\end{equation}
\begin{equation}
-k_x\left(Z_y^\prime(1)-(k_\perp^2+2\lambda)\int Z_y(1) \right)
+k_y\left(Z_x^\prime(1)-(k_\perp^2-2\lambda)\int Z_x(1)\right)=0
\label{eq:elec2}
\end{equation}
where $\int Z_i(z)=\delta a_i \exp(z/\delta)-\delta b_i \exp(-z/\delta)-ic_i/\kappa_z \exp(i\kappa_z z)+id_i/\kappa_z \exp(-i\kappa_z z)$.
%
\subsection{Squire and Orr-Sommerfeld modes}
\label{sec:sosmodes}
We shall now use the boundary conditions (\ref{eq:imp1}), (\ref{eq:imp2}), 
(\ref{eq:cont1}),(\ref{eq:cont2}),(\ref{eq:elec1}) and  (\ref{eq:elec2}) to find the sequences of values
 of $\delta$, $\kappa_z$, $\lambda$ as well as the properties of the related 
Eigenmodes.
Equations (\ref{eq:imp1}),(\ref{eq:imp2}),(\ref{eq:cont1}) and (\ref{eq:cont2}) impose 4 homogeneous conditions on the 
4 coefficients $A_z^1$, $A_z^2$, $A_z^3$ and $A_z^4$  in the expression of $Z_z$. This means that the eigenmodes of the dissipation operator can be divided 
into two categories.\\

1) For the modes of the first category, conditions (\ref{eq:imp1}),(\ref{eq:imp2}),(\ref{eq:cont1}) and (\ref{eq:cont2}) are redundant so that $Z_z\neq 0$. 
In this case, the determinant formed by those 4 conditions must be zero, which 
yields:
\begin{eqnarray}
4\frac{\kappa_z}{\delta}+\cosh(\kappa_z+1/\delta)(\kappa_z-1/\delta)^2 
\nonumber\\
-\cosh(\kappa_z-1/\delta)(\kappa_z+1/\delta)^2=0
\label{eq:orrsomm}
\end{eqnarray}
The sequence of values of $\kappa_z$ and $\delta$  is
determined by writing that $-\kappa_z^2$ and $\delta^2$ must both be roots of 
the dispersion equation (\ref{eq:disp}) for given values of $Ha, \lambda,
k_x$ and $k_y$. Eliminating $\lambda$ between (\ref{eq:delta}) and (\ref{eq:kappa_z}) yields the searched condition:
\begin{equation}
\frac{\kappa_z^2}{\delta^2}-k_\perp^2(k_\perp^2-\frac{1}{\delta^2}+\kappa_z^2 
+2Ha^2) =0
\label{eq:diss_kd}
\end{equation}
The system formed by (\ref{eq:orrsomm}) and (\ref{eq:diss_kd}) admits exactly 
one solution for $(\kappa_z,\delta)$ in each interval 
$[p\pi/2,(p+1)\pi/2]\times ]0,\infty[$ with $p\in \mathbb Z-\{0\}$. This 
 defines the sequence of possible values of $\kappa_z$ and $\delta$ for given 
$Ha$, $\lambda$, and $k_\perp$.  The related 
eigenmodes  have a non zero component along $\mathbf e_z$. By analogy with 
the eigenvalue problems which arise in the linear stability theory (see for instance \cite{drazin95}), we shall 
call them \textit{Orr-Sommerfeld} modes.\\

2) For the modes from the second category, conditions (\ref{eq:elec1}),
(\ref{eq:elec2}),(\ref{eq:cont1}) and (\ref{eq:cont2}) are not redundant so 
that $Z_z=0$. The continuity equation (\ref{eq:continuity}) then imposes that
 $k_xZ_x=-k_yZ_y$ so that the 
eigenmodes of this category are only determined by the 4 coefficients 
$A_x^1$, $A_x^2$, $A_x^3$, and $A_x^4$ (or alternately $A_y^1$, $A_y^2$, $A_y^3$, 
$A_y^4$).  $\delta$ and $\kappa_z$ are then determined using the electric 
conditions at the walls (\ref{eq:imp1}) and (\ref{eq:imp2}) which together with 
(\ref{eq:elec1}) and (\ref{eq:elec2}) impose 4 homogeneous conditions on the
4 coefficients $A_x^1$, $A_x^2$, $A_x^3$ and $A_x^4$ : the determinant 
formed by these conditions has to be zero for the eigenmodes not to be trivial:
\begin{eqnarray}
\left((\kappa_z^2-1/\delta^2)(k_\perp^4-\kappa_z^2/\delta^2)+4\kappa_z^2/\delta^2 \right)\sin(\kappa_z) \sinh(1/\delta) \nonumber\\
+4Ha^2\kappa_z/\delta(\cos(\kappa_z) \cosh(1/\delta)-1)=0
\label{eq:squire}
\end{eqnarray}
Here the system formed by (\ref{eq:squire}) and (\ref{eq:diss_kd}) admits 
exactly one solution for $(\kappa_z,\delta)$ in each interval $[p\pi/2,(p+1)\pi/2]\times ]0,\infty[$ 
with $p\in \mathbb Z$. The related eigenmodes have no velocity component along
$\mathbf e_z$ and those with $\kappa_z<\pi/2 $ exhibit only a weak dependence on
$z$ so that we may identify them as quasi two-dimensional modes. Following our 
analogy with linear 
stability problems, we shall name the modes of this second category
 \textit{Squire} modes.\\
These modes can be explicated by finding the three independent coefficients 
from $A_x^1$, $A_x^2$, $A_x^3$, and $A_x^4$. This is done by solving the system 
formed by  (\ref{eq:imp1}), (\ref{eq:imp2}),(\ref{eq:cont1}) and 
(\ref{eq:cont2}), which is at most of rank 3. $Z_y$ is obtained by 
$k_y Z_y=-k_x Z_x$ and $Z_z=0$. The associated eigenspace has therefore a 
dimension between 1 and 3.\\

In summary, for a given value of $Ha$, the eigenvalues 
$\lambda(k_\perp,\kappa_z)$  of the operator $\mathcal D_{Ha}$ are given by 
the dispersion equation (\ref{eq:diss_kd}). For a fixed value of $k_\perp$,
$\kappa_z$ takes one value in $]0, \pi/2[$ and 2 values in each interval 
$]p\pi/2,(p+1)\pi/2[$ with $p \in \mathbb Z-\{0\}$. This defines a discrete 
sequence of values for $\lambda$, which once sorted by increasing module yields 
the sequence of the least dissipative eigenmodes of the dissipation operator 
$\mathcal D_{Ha}$. At this point, the eigenmodes of the dissipation operator 
already exhibit a quite remarkable similarity with the known properties of
a flow between 2 parallel plates as each of those modes exhibits "core" velocity
 fluctuations of wavelength $(k_x,k_y,\kappa_z)$ as well as exponential 
 boundary layers of thickness $\delta$ along the walls at $z=-1$ and $z=1$.

\subsection{Numerical method}
\label{sec:3D_num}
By sorting the values of $\lambda(k_\perp,\kappa_z)$, we are now able to find 
the minimum of  $|\langle\Tr\mathcal D_{Ha}P_n\rangle|$ for any given value 
of $n$. 
We shall now perform this task numerically since the values of $\kappa_z$ 
cannot be found analytically. An upper bound for the attractor dimension as a 
function of $Ha$ and $Re$ then follows immediately. The numerical method is 
the same as the one already used in the case of a box with $2\pi$-periodic 
boundary conditions 
in the 3 directions of space 
\cite{pa03}: it consists of sweeping the $(k_\perp,\kappa_z)$ 
plane with the iso-dissipation (iso-$\lambda$) curves, starting at 
$|\lambda|=0$. For each value of $\lambda$, we count the number of $(k_\perp,\kappa_z)$ points enclosed in the corresponding iso-$\lambda$ curve and calculate
the sum of $\lambda$ over all of those points. Each point corresponds to an 
eigenvalue and should therefore be weighed by its multiplicity both when 
counting those points and building this sum. However it is simpler to 
build the sum by taking all multiplicities equal to one. The obtained sum for 
a given value of 
$n$ is a slightly looser lower bound for 
$| \langle \Tr \mathcal D_{Ha}P_n \rangle |$, but this 
may only affect the numerical constants appearing in the final results, and not
the dependence on $Ha$, $n$ and $Re$.\\
The difference between the present case with walls at $z=-1$ and $z=1$ and the 
periodic case is that  the sequence of real values of $\kappa_z$ replaces that 
of $k_z$ in the periodic case, which has integer values. 
It is important to notice that  the only imprecision brought by the numerical 
process is that of the truncation on the sequence of numerical values of 
$\kappa_z$ and $\lambda$, as well as that which results from the resolution of 
the systems of equations $\{(\ref{eq:orrsomm}),(\ref{eq:diss_kd})\}$ and 
$\{(\ref{eq:squire}),(\ref{eq:diss_kd})\}$ using the Newton method. 
However, since each solution of the system is bracketed, there is no risk of 
finding any spurious eigenvalues. In this sense, the method can be considered 
to give an exact result. It should also be pointed out that contrarily to the 
periodic case where $k_z$ has integer values, finding the sequence of values of 
$\kappa_z$ requires 
high precision arithmetics since the numbers involved in (\ref{eq:orrsomm}) and 
(\ref{eq:squire}) are very high, whereas the values of $\kappa_z$ are not. 
Using the LONG DOUBLE type in C language allows us to  reach values of $Ha$ up 
to $5000$ and $10^5$ modes. Computing the sequence of minimal modes for
 higher values of $Ha$ and $n$ would require the use of multi-precision 
libraries.\\
For given values of $n$ and $Ha$, the numerical algorithm yields directly the 
minimum of $|\langle \mathcal D_{Ha} P_n \rangle |$, as well as  $k_{\perp_m}$, 
$\kappa_{z_m}$, $\delta_{min}$, $\delta_{max}$, which are respectively the 
maximum values of $k_{\perp}$, $\kappa_z$, $\delta$, and the minimum value of 
$\delta$ over the set $n$ least dissipative modes.
The value of $n$ for which the trace of the total evolution 
operator is zero (\textit{i.e.} $\langle\Tr((-\mathcal D_{Ha}+ \mathcal B(.,\mathbf u)) P_n)\rangle=0)$ is an upper bound for the attractor dimension. Knowing 
the minimum of $|\langle \mathcal D_{Ha} P_n \rangle |$ and using 
(\ref{eq:tra}), the value of $Re$ for which $n$ is an upper bound for $d_M$ is 
given by $Re=\sqrt{2|\langle \Tr(\mathcal D_{Ha}P_n)\rangle |/n}$. The results 
 are plotted on figure \ref{fig:dimat}.

\section{Dimension and modes of 3D MHD attractors}
\label{sec:dim3D}

\subsection{Set of the least dissipative modes and attractor dimension}
\label{sec:3D}
\begin{figure}
\centering
\includegraphics[width=8.5cm]{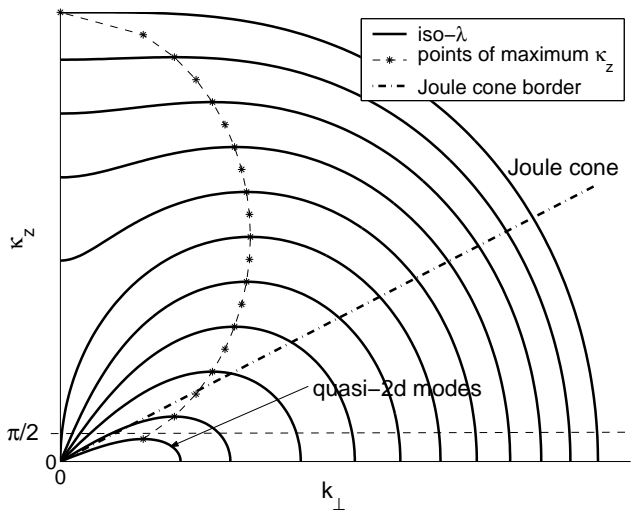}
\caption{Iso-dissipation curves in the Fourier space (one point represents all 
the eigenmodes with the same $(k_\perp,\kappa_z)$). Each curve encloses the set 
of the $n$ least dissipative modes of a given $n$. The properties (anisotropy 
and size of the smallest scales) from any combination of modes taken within the 
corresponding set can be 
read from the shape of the iso-$\lambda_m$ curve corresponding to the value of 
$n$ relating to the considered flow.
}
\label{fig:iso-lambda}
\end{figure}
\begin{figure}
\centering
\includegraphics[width=8.5cm]{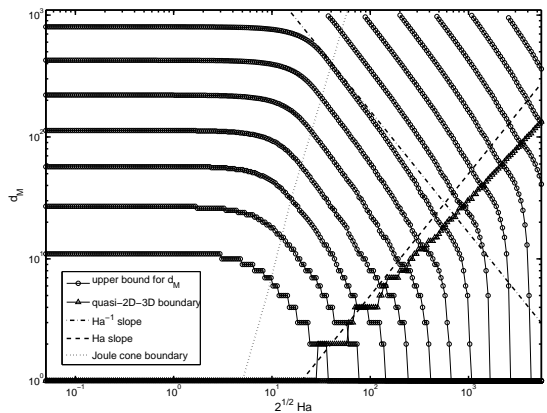}
\caption{Attractor dimension as a function of $Ha$ for fixed values of $Re$. 
3 regions appear which correspond to 3 types of sets of modes: 3D 
quasi-isotropic, 3D 
anisotropic, and quasi-2D. Note that the attractor dimension is strongly 
over-estimated in the quasi-2D case as the upper bound (\ref{eq:tra}) applies 
to all flows (3D). It can be improved in the case of quasi-2D flows as shown in 
section \ref{sec:q2D}}
\label{fig:dimat}
\end{figure}
Because the function $\lambda(k_\perp,\kappa_z,Ha \sqrt{2})$ has the same expression as the 
function $1/2\lambda(k_\perp,k_z,Ha)$ in the case with periodic boundary 
conditions 
in the 3 directions of space \cite{pa03}, the upper bound for the attractor 
dimension 
$d_M(Ha,Re)$ in both cases have similar behaviour. As in the periodic 
case, the maximum 
values  $k_{\perp_m}$ and $\kappa_{z_m}$  over the set of the $n$ least 
dissipative modes of $k_\perp$ and $\kappa_z$  can be 
deducted from the geometrical shape from the $\lambda(k_\perp,k_z)=\lambda_m$ 
curve, where $\lambda_m$ is associated to the most dissipative mode of the set
 (see figure \ref{fig:iso-lambda}). 
 
%
Increasing progressively the values of $Ha$ for a given value of $Re$, the 
upper bound for the attractor dimension exhibits three different behaviours 
which correspond to a 3D isotropic set of modes, a 3D anisotropic set and 
quasi-2D set respectively. For low values of $Ha$, $d_M$ varies little 
with $Ha$ and
 the set of eigenmodes of the dissipation which achieve the upper bound is 
 located in a circle centred on the origin in the $(k_\perp,\kappa_z)$ 
 plane. If one imagines a flow represented by a combination of such modes, it 
would be 3D and nearly 
 isotropic. For higher values of $Ha$, $d_M \sim Ha^{-1}$, just like in the 
case with periodic boundary conditions. The least dissipative modes are located 
 within a cardioid-shaped curve, with no modes in the vicinity of the 
 $\kappa_z$ axis. A flow featuring such modes would be  strongly anisotropic
 with vortices 
 elongated in the $z$ direction. The boundary between those first two types of
 sets of modes can be characterised by the disappearance (\textit{resp.} 
appearance) 
 of the last (\textit{resp.} first) mode of the form 
$(k_\perp=0,\kappa_z \neq 0)$. 
 When no such modes exist, the tangent at the origin of the iso-$\lambda$ curve
 corresponding to the most dissipative set of eigenmodes which achieve the 
 upper bound for $d_M$  can be defined. All modes are then located under this
 line which matches the concept of "Joule Cone" described by \cite{sm82}: in 
 the existing theories of turbulence, the energy-containing modes are expelled 
 from this cone of axis $(0,k_z)$ in the Fourier space for high values of 
$Ha$.\\
  The present case with walls at $z=-1$ and $z=1$ differs from the periodic 
case for higher 
 values of $Ha$, for which the related set of minimal eigenmodes of the 
 dissipation is quasi two-dimensional. In the periodic case, those modes all 
 satisfy  $k_z=0$ so that they are strictly two-dimensional and produce no
  Joule dissipation. In this case, $d_M$ doesn't  vary  anymore when $Ha$ is
 increased.
    In the case  with walls, the quasi two dimensional modes are  Squire modes 
    with  $0<\kappa_z<\pi/2$ which exhibit an  exponential velocity profile in the vicinity of the wall.  This means that the associated 
 values of $|\lambda|$ are higher in the case with  walls. When the least 
 dissipative modes are all quasi-2D Squire modes, some dissipation still arises 
 because of the  presence of the boundary layer profile at $z=1$ and $z=-1$. 
 This results in 
 $d_M$ decreasing rapidly as $Ha$ increases, contrarily to the periodic case. 
 The boundary layer properties of the least dissipative modes are more 
specifically studied  in  section \ref{sec:layers}. Also, since the upper bound 
for the trace of the 
 operator associated to the inertial terms (\ref{eq:tra}) is a general one which
 relies on no particular assumption
  on the flow's three or quasi two-dimensionality, it is unrealistic 
  for two-dimensional flows. An improved upper  bound for the 
  attractor dimension in the quasi-2D case is therefore derived in section 
  \ref{sec:q2D}, using a quasi two-dimensional model.
\subsection{Definition of the different boundaries separating classes of least 
dissipative modes}
For a given value of $Ha$, different types of sets of least dissipative modes 
are encountered when $n$ (or $Re$) increases. For the sake of clarity, we shall 
summarise here the 3 boundaries separating those different types of modes and 
their precise definitions:
\begin{itemize}
\item \textit{Boundary between  quasi-2D and 3D sets of modes:}\\
Squire modes such that $\kappa_z<\pi/2$ are the least dissipative of all and 
represent quasi-2D velocity fields. A set of modes containing only such modes 
is quasi-2D. For a given value of $Ha$, the set of the least dissipative modes 
crosses the quasi-2D- 3D boundary either when the first Orr-Sommerfeld 
mode or the first squire mode with $\kappa_z>\pi/2$ appears.
\item \textit{''Joule cone'' boundary}\\
For strong values of $Ha$, the set of least dissipative modes contains no mode
located along the $(O,\kappa_z)$ axis so that all modes are located outside of 
the Joule cone, defined by the tangent at $(k_\perp,\kappa_z)=(0,0)$ to the set 
of least dissipative modes (see figure \ref{fig:iso-lambda}). No Joule cone can therefore be defined beyond the value of 
$n$ at which the first mode of the form $(0,0,\kappa_z)$ appears within the 
set. This defines what we call the \textit{Joule cone boundary}, which 
separates a strongly anisotropic (with Joule cone) from a weakly 
anisotropic MHD set of modes (without Joule cone).
\item \textit{Boundary between sets with single and sets with multiple 
boundary layer thickness}\\
The evolution of the set of values of $\delta$, reached within the set of the 
least dissipative modes when $n$ increases, exhibits a sharp transition between 
sets for which all values are very close to one particular value (for low $n$)
 and sets with more important scattering (larger $n$). The first one corresponds
 to a laminar boundary layer state with a well defined thickness, whereas the 
 second exhibits strong similarities with the turbulent Hartmann boundary layer
 as more quantitatively explained in section \ref{sec:layers}. For a given 
 value of $Ha$, we define this  transition as the intersection between
the line $\sqrt{2}\delta Ha=1$ and the asymptote to the 
curve $\delta_{min}(n)$ when $n\rightarrow\infty$ (see figure \ref{fig:delta}).
\end{itemize}
When $Ha$ is increased from $0$ for a given value of $n$ or $Re$, these 
boundaries are encountered in the reverse order. At this point it is important
to underline that the properties mentioned in this section are those of the 
least dissipative modes and not of the real flow. The link between the two is
discussed in conclusion.\\
\subsection{Analytic approximation for the upper bound for $d_M$}
We shall now derive some analytical estimates for the attractor dimension 
so as to be able to further compare our results to the results for the size 
of the dissipative scales available from \cite{pa03}. We consider separately
either side of the Joule cone boundary defined in the previous section, 
\textit{i.e.} 3D anisotropic sets of modes  (with Joule Cone) and weakly 
anisotropic 3D sets (without Joule cone).\\

As the modes are spread uniformly in the Fourier space, with $1/\pi^3$ modes 
per unit of volume, the value of $\lambda_m$  can be found by writing that in 
the $1/8^{th}$ space $k_x>0, k_y>0, \kappa_z>0$, the volume enclosed by the 
iso-$\lambda_m$ curve should be $n/(8\pi^3)$ 
. For sufficiently high values of $\kappa_{z_m}$ and $k_{\perp_m}$, 
this can be expressed using integrals:
\begin{equation}
8\pi^3\int_{V_{\lambda_m}}dk_xdk_ydk_z=n
\label{eq:nmodes}
\end{equation}
The trace of $\mathcal D_{Ha} P_n$ similarly expresses as:
\begin{equation}
\Tr(\mathcal D_{Ha} P_n)=8\pi^3\int_{V_{\lambda_m}}\lambda(k_x,k_y,k_z)dk_xdk_ydk_z
\label{eq:trdint}
\end{equation}

The set of $n$ least dissipative modes in the Fourier space is enclosed in the 
same iso-$\lambda$ curve as the set of $8n/\pi^3$ least dissipative modes in the
$2\pi-$periodic case. The attractor dimension and the corresponding values of 
$k_{\perp_m}$ are then found by replacing $n$ with $8n/\pi^3$ in the results from 
study with periodic walls (\cite{pa03}, p3176). Two distinct cases are found:\\

1) \emph{Anisotropic sets of modes (with a Joule cone)}\\
For 3D anisotropic  set of least dissipative modes, all modes are located 
within an elongated cardioid with no mode of the form $k_\perp=0$. The Joule 
cone of axis $(O\kappa_z)$ exits and its half-angle it that of the tangent at 
the origin of the iso-$\lambda_m$ curve (see figure \ref{fig:iso-lambda}):
 \begin{equation}
 \sin\theta_m=\sqrt{\lambda_m/Ha^2} =2^{-1/8}2\pi^{-5/4}n^{1/4}Ha^{-3/4}%
 \label{eq:cone_angle}
 \end{equation}
 All modes are located outside of this
 cone in the Fourier space. In this case, the upper bound for the attractor 
 dimension is:
 \begin{equation}
 d_M \leq \frac{9}{256\pi\sqrt{2}}\frac{Re^4}{Ha}
 \label{eq:dimat1}
 \end{equation}
 and the related bounds for $k_{\perp_m}$ and $\kappa_{z_m}$  are:
 \begin{eqnarray}
 k_{\perp_m}\leq \left(\frac{3}{2\pi^8}\right)^{1/4}  Re
 \label{eq:kpmd1}\\
 k_{z_m} \leq \frac{1}{\sqrt{2}}\left(\frac{3}{2\pi^8}\right)^{1/2}\frac{Re^2}{Ha}
 \label{eq:kzmd1}
 \end{eqnarray}
  These sets of modes describe are 3D anisotropic. For a 3D anisotropic flow, 
 heuristic considerations of the Kolmogorov type predict $d_M\sim Re^2/Ha$, 
$k_{\perp_m}\sim Re^{1/2}$ and $k_{z_m}\sim Re/Ha$ (see \cite{pa03}).\\

2) \emph{Weakly anisotropic sets of modes (without Joule cone)}\\
In the case where no Joule cone can be defined  within the set of 
dissipative mode, the related set is quasi-isotropic with:
\begin{equation}
d_M \leq \frac{5\sqrt{30}}{216\sqrt{\pi}} Re^3 \left(1-\frac{4\pi^3}{3}\frac{Ha^2}{Re^2}
\right)^{3/2}
\label{eq:dimat2}
\end{equation}
and corresponding $k_{z_m}$ and $\kappa_{z_m}$: 
\begin{eqnarray}
k_{\perp_m}=\sqrt{\frac{5}{\pi}}Re\left(1-2\pi^3\left(\frac{2}{9}+\frac{1}{15\pi^2}\right)\frac{Ha^2}{Re^2}\right)^{1/2}
\label{eq:kpmd2}\\
\kappa_{z_m}=\sqrt{\frac{5}{\pi}}Re\left(1-2\pi^3\left(\frac{2}{9}+\frac{4}{15\pi^2}\right)\frac{Ha^2}{Re^2} \right)^{1/2}
\label{eq:kzmd2}
\end{eqnarray}
In the limits of small Hartmann numbers, (\ref{eq:dimat2}) recovers the 
classical hydrodynamic bound for 3D turbulence of $Re^3$ (see for instance 
\cite{constantin85_jfm}). 
The boundary between the 3D weakly anisotropic and 3D anisotropic sets of 
modes can be characterised by the appearance of the Joule cone. This happens 
when $\sin \theta_m=1$, \textit{i.e.}:
\begin{equation}
\frac{Ha}{Re}=\sqrt{3\pi^3}2^{5/4}
\label{eq:cone}
\end{equation}
As in the periodic case, the bounds for $d_M$, $k_\perp$ and $\kappa_{z_m}$
 exhibit the same dependence on $Ha$ as the heuristic predictions which 
suggests 
that our estimate renders the effect of the Lorentz force on the small 
scales realistically. The powers of $Re$ are however overestimated, but this 
can be inferred to the estimate for the expansion rate of the inertial terms 
used in (\ref{eq:tra}) which is known to be too high. The problem of finding an 
 optimal estimate for these terms is 
to this day still open.\\ 
Also, one expects that when the lorentz force becomes of the order of magnitude 
of inertial effects in the real flow (\textit{i.e.} $Ha^2\gtrsim Re$), the 
vortices start stretching in the direction of the magnetic field, so fewer of 
them should be present. Estimates from Kolmogorov-like arguments imply indeed 
that for a given Reynold number, the number of vortices
in the flow without magnetic field ($\sim Re^{3/4}$) is larger than that in a 
flow with strong magnetic field ($\sim Re/Ha$)as soon as $Ha \gtrsim Re^{-1/4}$.
 This property is recovered from the results on the upper bound for 
$d_M$ since $d_M(Ha)$ clearly decreases when $Ha$ increases at any fixed $Re$ 
(see figure \ref{fig:dimat}). (\ref{eq:dimat1}) and (\ref{eq:dimat2}) however 
imply that this decrease starts for $Ha>Re$, which in most real cases would 
correspond to a laminar flow (\textit{i.e.} $d_M=0$ !). This is clearly 
unrealistic and stems again from the fact that the estimate for the trace of 
the inertial terms $nRe^2$ from which $d_M$ is derived  for both 
hydrodynamic and strongly anisotropic MHD flows is even further than the 
heuristic estimate derived 
using Kolmorogov-like arguments in the MHD case ($\sim n Re$) than in the 
hydrodynamic case ($\sim n Re^{3/2}$).

In the two cases  of 3D flow with and without Joule cone, the upper bound for 
the attractor dimension exhibits the same dependence on the non-dimensional 
numbers as in the case with periodic boundary conditions. This confirms the 
tendency already observed on the numerical results from the previous section. 
It also suggests that problems
with periodic boundary conditions in the 3 directions of space provide some 
scaling laws which are relevant to cases involving more realistic 
boundary conditions like impermeable walls as long as the velocity field is 
strongly 
three-dimensional. This relevance however breaks down for quasi two-dimensional 
velocity fields, the dynamics of which is controlled by the boundary layers 
which arise along the walls perpendicular to the magnetic field.\\

\subsection{Boundary layer properties}
\label{sec:layers}

\begin{figure}
\centering
\includegraphics[width=8.5cm]{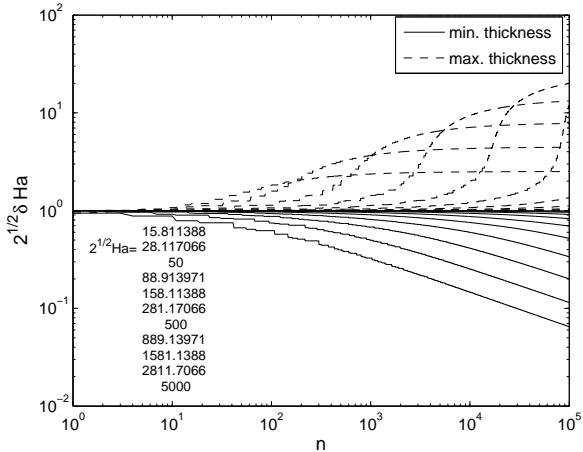}
\caption{Maximum and minimum of the boundary layer thickness $\delta$ for $n$ 
the set of modes
which minimise the dissipation, as a function of $n$, for different values of 
$Ha$. For each $Ha$, a boundary appears between sets of modes where all modes 
have a boundary layer thickness close to $1/Ha$ (\textit{i.e.} 
$\delta_{min}\simeq\delta{max}$) and sets for which a broader spectrum of 
values of $\delta$ is present (\textit{i.e.} $\delta_{min}<<\delta{max}$).
 The value of $n$ at which this boundary is found place increases 
with $Ha$.}
\label{fig:delta}
\end{figure}

\begin{figure}
\centering
\includegraphics[width=8.5cm]{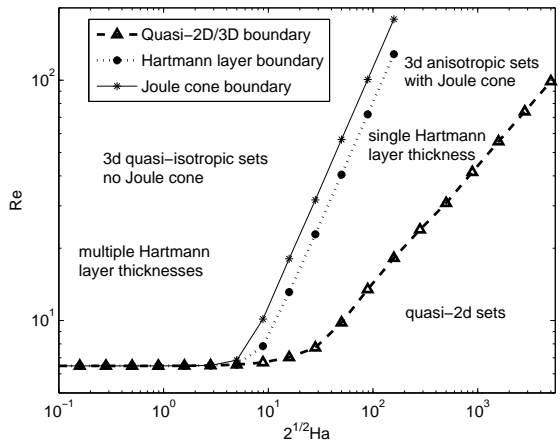}
\caption{Map of the different types of sets of modes in the $(Ha,Re)$ plane}
\label{fig:trans_turbu}
\end{figure}

We now turn our attention to the influence of the walls on the modes minimising 
 the dissipation, as there lies the most important difference between the  
periodic case and the case with walls. More precisely, we shall study
 the values of the boundary layer thickness $\delta$ which characterises 
the eigenmodes of the dissipation.\\
As mentioned in section \ref{sec:eigen}, each mode $(k_x,k_y,\kappa_z)$ can 
alternately  be represented by the triplet $(k_x,k_y,\delta)$. Figure 
\ref{fig:delta} represents the evolution of the minimum (resp. maximum) value 
of $\delta$ reached within the set of the $n$ least dissipative modes 
$\delta_{min}$ (resp. $\delta_{max}$) as a function of $n$, obtained 
numerically from $\{(\ref{eq:orrsomm}),(\ref{eq:diss_kd})\}$ and
$\{(\ref{eq:squire}),(\ref{eq:diss_kd})\}$. For each fixed value of $Ha$, 
 there is an approximate  value $n_H$ of $n$, which marks a boundary 
between two types of sets of least dissipative modes:
 for $n<n_H$, all modes are characterised by a value
of $\delta$ close to $1/Ha$ (\textit{i.e.} $\delta_{min}\simeq\delta{max}$)
whereas for $n>n_H$, the set of least 
dissipative modes exhibits a much broader spectrum of values of $\delta$ 
(\textit{i.e.} $\delta_{min}<<\delta{max}$). A 
 velocity field represented by a combination of modes taken below this 
value of $n$ would exhibit a laminar boundary layer of thickness $1/Ha$, with 
an exponential 
profile.  This matches exactly the prediction of the laminar Hartmann layer 
theory (see for instance \cite{moreau90}). For $n>n_H$, 
some modes appear with a thicker boundary layer, as well as modes with a 
layer much thinner than $1/Ha$. 
This two-layer structure strongly resembles that of the 
theoretical prediction of \cite{albouss00} for the turbulent Hartmann layer, 
which also involves such a double deck structure, with a viscous sublayer.\\
In order to quantify how $n_H$  depends on the parameters 
 $Ha$, $n$ and $Re$, we define it quantitatively as the value of $n$ where   
the line $\delta Ha=1$ (see \ref{fig:delta}) intersects the asymptote to the 
curve $\delta_{min}(n)$ for each value of Hartmann and we have plotted the 
result on figure 
\ref{fig:trans_turbu}. It is   remarkable that the boundary between sets of 
modes with a single boundary layer thickness and sets with multiple boundary
layer thickness is never found between quasi-2D sets of modes, but always where
3D modes are present.   This strong property seems to differ from  
  the results of \cite{krasnov04} who studied numerically the transition to 
  turbulence of the Hartmann layer in channel flows and found that there can 
  exist a  two-dimensional region outside of the turbulent Hartmann layer. 
A comparison to the real flow in this particular instance is however not 
straightforward since 
even if we assume that the 
flow can indeed be represented by a combination of least dissipative modes 
some of which are 3D, 
the flow can still exhibit some 2D features, as explained in conclusion.\\
These numerical results can be supported by some analytical expressions for the 
minimum and maximum boundary layer thickness, which can be
obtained by eliminating $k_\perp^2$ between (\ref{eq:kappa_z}) and 
(\ref{eq:delta}). This defines the iso-$\lambda$ curves in the 
$(\delta,\kappa_z)$ plane, which
enclose the set of least dissipative modes. Then remarking that the maximum and 
minimum values of $\delta$ are obtained for $\kappa_z=0$ yields the expressions
for $\delta_{min}$ and $\delta_{max}$ \footnote{$\kappa_z$ is never zero so these values of $\delta_{min}$ and $\delta_{max}$ are never reached: $\delta_{max}$
is a close upper bound and $\delta_{min}$ is a close lower bound.}:
\begin{eqnarray}
\delta_{min}=\frac{1}{\sqrt{2}}\frac{1}{\sqrt{Ha^2-\lambda_m}}
\label{eq:dmin}\\
\delta_{max}=\frac{1}{\sqrt{2}}\frac{1}{\sqrt{Ha^2+\lambda_m}}
\label{eq:dmax}
\end{eqnarray}
%
In the case of a 3D set of modes with Joule cone 
($1/\sqrt{2}Ha \leq \lambda_m \leq Ha^2$), $\delta_{min}$ and $\delta_{max}$ are 
expressed as functions of $Ha$ and $Re$ using (\ref{eq:cone_angle}) and 
(\ref{eq:dimat1}):
\begin{eqnarray}
\delta_{min}=\frac{1}{\sqrt{2}}\frac{1}{Ha\sqrt{1+\frac{1}{2\pi^3}\frac{Re^2}{Ha^2}}}
\label{eq:dmin_re}\\
\delta_{max}=\frac{1}{\sqrt{2}}\frac{1}{Ha\sqrt{1-\frac{1}{2\pi^3}\frac{Re^2}{Ha^2}}}
\label{eq:dmax_re}
\end{eqnarray}
For low Reynolds numbers (or low values of $|\lambda_m|$ in (\ref{eq:dmin}) 
and (\ref{eq:dmax})), $\delta_{min}$ and $\delta_{max}$ are very close to 
each other, whereas they separate quickly when $\frac{Re}{Ha}$ 
approaches unity, as confirmed by the numerical results on figure 
\ref{fig:trans_turbu}. Remarkably, $\frac{Re}{Ha}\sim 1$ coincides also 
approximately with the disappearance of the Joule cone (\ref{eq:cone}).\\
As for the turbulent quantities discussed in section \ref{sec:3D}, the  
boundary layer properties of the least dissipative modes exhibit some
striking similarities with that of existing theories 
\cite{moreau90}. Quantitatively, 
the boundary layer thickness associated to those modes has the same 
dependence on $Ha$ as 
that of real Hartmann layers. 
This adds up to the conclusion of \cite{pa03} who found that the set of least 
dissipative modes in a 3D periodic box share many of their properties with those of the real turbulent low (Joule cone angle, small scales, boundary between 
quasi-2D and 3D sets of modes). 
Indeed, when walls are present,  not only are the turbulent properties of the 
core flow recovered, but also some of the fine properties of the Hartmann 
boundary layers are closely mimicked by the least dissipative modes (thickness,
boundary between sets of modes with a single boundary layer thickness of $Ha^{-1}$ and sets of modes with  multiple thickness.)

\section{Attractor dimension for a 2D MHD model}
\label{sec:q2D}

To conclude this search of an upper bound for the attractor dimension of a 
turbulent MHD flow in a box with Hartmann walls, we shall now derive a tighter
bound in the case where the flow is two-dimensional by using 2D 
motion equations. Such equations are often used to model quasi two-dimensional 
flows between two parallel insulating planes. They are obtained by averaging 
the 
full 3D Navier-Stokes equations along the direction of the magnetic field 
between the two planes. The model is closed by assuming a particular velocity
profile along this direction. \cite{sm82} were the first to propose such 
a model, in which this particular velocity profile was derived from first order 
matched asymptotics using the small parameters $1/Ha$ and $1/N$. At this order,
the velocity profile does not vary along the direction of the magnetic field, 
except in the vicinity of the walls where Hartmann boundary layers of thickness 
$1/Ha$ 
develop with an exponential velocity profile. For a distance $2L$ 
between the Hartmann walls, the resulting model can
be written as follows in dimensionless terms:
\begin{equation}
\frac{\partial}{\partial t}\mathbf u + \mathbf u \cdot \mathbf \nabla \mathbf u
+\nabla p= \mathbf \nabla ^2 \mathbf u - H\!a
\mathbf u+ {\mathcal G} \mathbf f
\label{eq:sm82}
\end{equation}
%
where the Hartmann number is defined as previously 
$H\!a=\sqrt{\sigma / (\rho \nu )} B L$ and the Grashoff number is the 
dimensionless measure of a two-dimensional forcing $\mathbf f$, defined 
as ${\mathcal G} = \vert \mathbf f \vert _{L^2} L^2 / \nu$. As in the previous 
sections, this forcing ensures that the flow is steady on average as the energy it injects in the flow is balanced by the dissipation  (here either by viscous 
friction of by the linear friction term).

Even though this model is obtained at the cost of an approximation on the 
full 3D equations, the properties of the related dynamical system are 
 still interesting
to investigate since this model has proved its accuracy in many instances 
\cite{frank01, psm00, psm05}. We shall now derive an upper bound for the attractor 
dimension $d_{2D}$ for the related 
problem with $L-$periodic boundary conditions in the two directions of space 
$\mathbf e_x$ and $\mathbf e_y$.\\
The only difference between (\ref{eq:sm82}) and the two-dimensional 
Navier-Stokes equation is the dissipation term, with related operator
$\mathcal D_{2D}=\nabla^2-H\!a\mathcal I$, where $\mathcal I$ is the identity.
Application of the method described in section \ref{sec:method} yields an 
upper bound 
for the growth rate of any $n-$dimensional volume located in the vicinity of 
the attractor:
\begin{equation}
\langle\Tr \mathcal A P_n\rangle = \langle\Tr \mathcal B(\cdot,\mathbf u)P_n +\Tr \mathcal D_{2D}P_n\rangle
\end{equation}
The trace of the Hartmann friction operator can be written exactly $\Tr ( -H\!a {\mathcal I} P_n) = - H\!a n$. The trace of the other operators is taken from \cite{constantin88}, equation (3.19), and the trace of the global operator takes the following form:
%
\begin{equation}
\langle\Tr \mathcal A P_n\rangle \leq \frac{\Tr ({\nabla ^2} P_n)}{2} + c C^{4/3}\left(1+log C)^{2/3}\right) - H\!a n 
\label{trace2D}
\end{equation}
where $C=L^2/(4\pi^2\nu)\sup_{\mathbf u}\langle |\nabla^2 \mathbf u|^2\rangle
^{1/2}$ and $c$ is a dimensionless constant of order unity.
As it can be shown that $\Tr ({\nabla ^2} P_n ) \leq - (8\pi^2) n^2$, the largest root of (\ref{trace2D}) 
provides an upper bound 
for the attractor dimension:
\begin{equation}
d_{2D} \leq -\frac{Ha}{2\pi} + \sqrt{\frac{Ha^2}{4\pi^2} +4cC^{4/3}(1+log C)^{2/3}}
\label{eq:dsm_c}
\end{equation}
Since \cite{ohkitani89} has shown that the estimate for the trace of the inertial term derived 
for the classical 2D theory was at least log-optimal, 
the final bound (\ref{eq:dsm_c}), can be reasonably expected to be just as 
optimal.\\

We now need to express $C$ in terms of the governing dimensionless numbers $H\!a$ and $\mathcal G$. This is done in the appendix (\ref{appendix}). For a large-scale forcing, the following relationship is derived:
\begin{equation}
C \leq \frac{\mathcal G}{2 \sqrt{H\!a}}
\label{eq:CvsGHa}
\end{equation}
Neglecting logarithmic corrections, equation (\ref{eq:dsm_c}) can then be written:
\begin{equation}
\frac{d_{2D}}{c'} \leq -Ha + \sqrt{Ha^2 +c'' {\mathcal G}^{4/3}H\!a^{-2/3} }
\label{eq:dsm_GHa}
\end{equation}
where $c'$ and $c''$ are other dimensionless constants of order unity. 

Two limit cases are of interest, depending on the relative strength of ${\mathcal G}$ and $H\!a^2$. When ${\mathcal G}$ is much larger than $H\!a^2$, the attractor dimension can be bound as ${d_{2D}} \leq  {\mathcal G}^{2/3}H\!a^{-1/3}$, where constants of order unity are dropped. On the contrary, when $H\!a^2$ is much larger than ${\mathcal G}$, an expansion of (\ref{eq:dsm_GHa}) leads to ${d_{2D}} \leq  {\mathcal G}^{4/3}H\!a^{-5/3}$. 

\begin{figure}
\begin{center}
\psfrag{G}{$\mathcal G$}
\includegraphics[width=8.5cm]{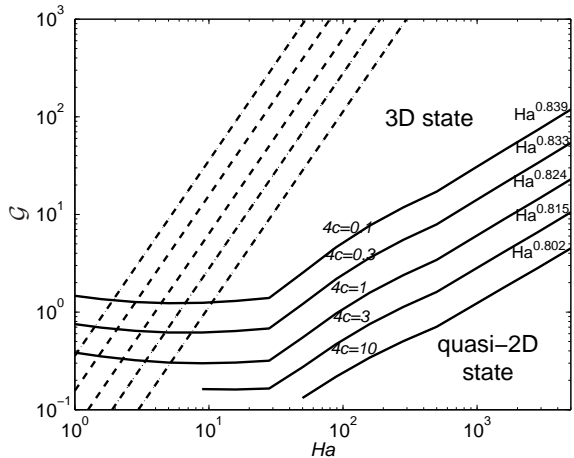}
\caption{Quasi-2D/3D boundary  in the ${\mathcal G}-H\!a$ plane for different 
values of the constant $4c$ appearing in equation (\ref{eq:dsm_c}) (solid lines). The 
dashed lines represent the according $\mathcal G=2H\!a^2/(4c\pi^2)^{3/4}$ 
condition for the two terms under the radical in (\ref{eq:dsm_c}) to be of the 
same order of magnitude. Asymptotic slopes to the transition curves obtained by
interpolation  are also given: they depend little on $c$ and are all around $Ha^{4/5}$.}
\label{fig:trans_c_ha}
\end{center}
\end{figure}

Finally, it is possible to use this 2D result on attractor dimension where inertial terms are probably well estimated to gain knowledge on the boundary between 
quasi-2D and 2D sets of modes. If we interpret $d_{2D}$ as the number of modes 
present in the quasi-2d set of modes which achieves the upper bound for $d_M$ for
 a given value of $Ha$ and $\mathcal G$, this quasi-2D/3D boundary obtained 
numerically in section \ref{sec:3D} as a function of
$n$ (number of modes) and $Ha$ can also be expressed in terms of ${\mathcal G}$ 
and $Ha$. The result is reported on figure \ref{fig:trans_c_ha}.
Since the bound (\ref{eq:dsm_GHa}) can reasonably be expected to be optimal, 
so can be the quasi-2D/3D boundary. The curve suggest that for sufficiently 
large Hartmann numbers, the quasi-2D/3D boundary is approximately located at 
$\mathcal G\sim Ha^{4/5}$. This prediction has the advantage to be comparable 
to the transition between quasi-2D and 3D turbulence that occur in experiments 
as the case with walls studied here corresponds to realistic experimental 
conditions. Also, for a given value of $Ha$, the quasi-2D/3D boundary in the 
case with walls happens for higher values of $\mathcal G$ as in the periodic case ($\mathcal G\sim Ha^{2/3}$ from \cite{pa03}). This is a consequence of the 
extra dissipation from the Hartmann layer friction, which is absent in the 
periodic case.

%

\section{Concluding remarks}
\label{sec:concl}
We have derived an upper bound for the fractal dimension of the attractor 
in the case of steady on average forced flow  between two electrically 
insulating parallel plates perpendicular to a strong steady homogeneous 
magnetic field and over the whole 
range of non-dimensional numbers $Ha$ and $Re$ in 3D or $Ha$ and $\mathcal G$ 
in the quasi-2D case. The derivation of this bound has been achieved by 
sorting and counting the eigenmodes of a linear operator which is very 
close to that representing the total dissipation (that is Joule and viscous).
Those modes strongly resemble the real turbulent flow in the sense that they
exhibit both a wavelength and a boundary layer profile in the vicinity of the 
walls. This lead us to define some boundaries between different sets of these 
modes: between quasi-2D  and  3D anisotropic sets (with a Joule cone),
 between sets with and without a Joule cone (strongly/weakly  
anisotropic) and also between sets with modes exhibiting boundary layer profiles  with a single boundary layer thickness of $1/Ha$ and those with boundary layer
profiles  spanning a broader spectrum of  thicknesses.\\
	The apparent similarity between the properties of the least dissipative
modes (wavelength, boundary layer thickness) and those of the real flow, as well
as between the boundaries separating sets of modes and transitions between 
turbulent states led us to compare the exact mathematical results obtained in 
this paper to the heuristics available for turbulent flows. Except for the transition to 
turbulence in the Hartmann layer, those transitions, lengthscales and
boundary layer scaling laws exhibit too high exponents for the Reynolds number 
while the exponents of the Hartmann number are correct. One reason is that the 
estimate for the inertial terms we start from is known to be an overestimate, 
whereas the estimation on the linear part of the evolution operator which 
involves $Ha$ is likely to be optimal. Interestingly, those scaling laws are 
also identical up to a constant to those derived for the case with periodic 
boundary conditions \cite{pa03}, when the flow is far enough from a 2D or 
quasi-2D state. This underlines that periodic boundary conditions are relevant 
to the description of 3D wall-bound flows, but not to that of quasi-2D wall 
bound flows.\\
The "boundary layer" part of the eigenmodes of the dissipation operator 
shows a sharp boundary between sets of modes with a single Hartmann layer of 
thickness $1/Ha$ and sets with multiple boundary layer thicknesses. The law 
$Ha/Re=constant$ for this boundary matches the heuristics known for the 
transition to turbulence of the Hartmann layer in the channel flow configuration
 on both the exponents of $Ha$ and $Re$. It should, however, be kept in 
mind that in channel flows, $Re$ is based on the average velocity whereas our 
Reynolds number is based on a maximum fluctuating velocity. Together with the 
finer description of the 
quasi-2D case and the transition between quasi-2D and 3D turbulence, those
"Hartmann layer" properties are the most significant novelty of this work, 
compared to the previous study with walls \cite{pa03}.\\
In the last part of this work, we have addressed the case of quasi-2D MHD 
turbulence by using a the 2D model from \cite{sm82}. The main advantage of this 
method is that optimal bounds for the 2D inertial terms are available from the 
literature. This has allowed us to derive both estimates on the size of small 
scales as well as a law for the boundary between quasi-2D and 3D sets of modes, 
expressed as a function of the Grashoff and the Hartmann numbers. The question 
of knowing whether those results are still optimal however remains open. The  
 answer could hopefully come from laboratory experiments, since the 
presence of Hartmann walls in the present study makes the results derived here 
directly comparable to experiments.\\

Lastly, the remarkable similarity between the least dissipative modes derived 
in this work and the real flow leads to the central question of knowing whether 
they do represent the flow or not. Answering this question is anything but 
trivial and the authors believe it might open some interesting perspectives in 
the study of anisotropic turbulent flows. We shall now finish this discussion by 
giving some ideas of what it involves.
The picture is seemingly ''simpler'' in hydrodynamic turbulence with periodic 
boundary conditions where the least dissipative mode are the usual isotropic 
sequence of Fourier modes used in spectral DNS. Those modes are a basis of 
$\mathcal L_2$ on which any solution including the ''real'' flow can therefore 
be expanded. Since it is mathematically proved that the number of vortices 
in the flow obtained from the Kolmogorov scales by a geometric argument 
matches the attractor dimension up to a multiplicative constant of order one
(\cite{constantin85_jfm}), any optimal upper bound for this dimension gives 
an exact account of the size of the small scales and tells exactly which 
finite basis of Fourier modes is required to fully describe the flow at a 
given $Re$. The full flow information including non-modal events and 
intermittency is, however, not contained in the basis of least dissipative 
modes itself, but in the time-dependent coefficients of the Fourier 
expansion, found, for example, by spectral DNS.\\
Now in the case of MHD turbulence, showing that the number of vortices obtained
 from the small scales heuristics is also the attractor dimension up to a 
multiplicative constant, poses no particular difficulty (but goes beyond the 
scope of this work). In contrast, showing that the least dissipative modes 
form a basis of $\mathcal L_2$
\footnote{In fact, the system of interest is rather the the biorthogonal set 
$\mathcal B_{\mathcal D_{Ha}}$ made of the canonical
system of the dissipation operator $\mathcal D_{Ha}$ and that of its adjoint
operator \cite{keldysh51,keldysh71}, but this technical difference,  due to the 
non-orthogonality of the $\mathcal D_{Ha}$ operator is of no influence for our 
present purpose}
 might require an elaborate mathematical proof. 
\cite{Ilin76} has shown that such a 
system associated to any ordinary differential operator of order $n$ formed a 
basis of $\mathcal L_2$, but this result has yet to be extended to partial 
differential operators. Let's assume for a moment that this property is valid, 
as in most of the non-mathematical literature. Then, as in hydrodynamic 
turbulence, the attractor dimension gives the size of the small scales and 
indicates which basis of least dissipative modes is required to fully describe 
the flow for a given values of $Ha$ and $Re$. Since the least dissipative modes 
contain not only the two wavelength $k_\perp$ and $\kappa_z$ but also the 
boundary layer thickness, this implies the following:
\begin{itemize}
\item If the set of least dissipative modes is 2D, so is the resulting flow
\item If the set of least dissipative modes has a Joule cone, so has the resulting flow
\item If the set of least dissipative modes contains only laminar boundary 
layer profiles of thickness $1/Ha$, then the Hartmann boundary layers present 
in the real flows are laminar. 
\end{itemize}

These results, like those for the hydrodynamic case  hold for the 
modes which achieve any upper bound for the attractor dimension. Now as in 
the hydrodynamic case, the information about non-modal 
events and intermittency is not contained in the basis of least dissipative 
modes itself, but in the coefficients of the expansion on this basis, which 
represent the real flow. This implies, for instance that even when the set of 
least dissipative modes contains multiple thicknesses, the Hartmann layer 
can still be laminar, at least intermittently if all the modes with $\delta\neq 0$ have near zero coefficients. The same remark applies to the quasi-2D/3D 
transition where for instance a flow represented by a 3D set of modes can be 
quasi-2D, at least intermittently.\\

	We are now left with two tasks: mathematicians have to try and prove 
(or disprove !) that $\mathcal B_{\mathcal D_{Ha}}$ is a basis of 
$\mathcal L_2$. Physicists may not wait for the mathematicians and already 
perform spectral DNS using these modes in order to calculate the real flow. This may turn out to be very cost-effective as the number of modes needed decreases as 
$1/Ha$, and also because the thin Hartmann layer is finely described by 
only very few of these modes. The description of this layer traditionally 
imposes a strong  limitation to such numerical calculations as the number of 
Tchebychev polynomials it requires increases rapidly with $Ha$. Up to now, this has held cases with $Ha=1000$ out of the reach of Direct Numerical Simulations.\\

The authors would like to acknowledge the partial financial support from the 
Leverhulme Trust, under Grant $F/09 452/A$.\\

\appendix

\section{Derivation of a bound on $C$ in terms of $\mathcal G$ and $H\!a$, in two-dimensional turbulence}
\label{appendix}
{
The curl of the governing equation (\ref{eq:sm82}) can be written:
\begin{equation}
\frac{\partial }{\partial t}{\mathbf \omega} + ({\bf u}\cdot{\bf \nabla}){\mathbf \omega} = - H\!a \ {\mathbf \omega} + {\bf \nabla}^2 {\mathbf \omega} + {\mathcal G} \ {\bf \nabla}\times{\bf f}, \label{Vortadim}
\end{equation}
where $\mathbf \omega$ is the vorticity. Multiplying the vorticity equation above by $\mathbf \omega$ and integrating over the two-dimensional domain yields the enstrophy equation:
\begin{equation}
{\bf 0} = - H\!a \ \int {\mathbf \omega}^2 dV + \int {\mathbf \omega} \cdot {\bf \nabla}^2 {\mathbf \omega} dV + {\mathcal G} \  \int {\mathbf \omega}\cdot {\bf \nabla}\times{\bf f}, \label{Enstrophy}
\end{equation}
Integrating by part the term in the middle allows us to write the equation under the following form:
\begin{equation}
H\!a  \int {\mathbf \omega}^2 dV + \int \left( {\bf \nabla}^2 {\bf u}\right) ^2 dV = {\mathcal G}   \int {\mathbf \omega}\cdot {\bf \nabla}\times{\bf f} dV. \label{Enstrophy2}
\end{equation}
Application of Cauchy-Schwartz inequality leads to:
\begin{equation}
H\!a  \int {\mathbf \omega}^2 dV + \int \left( {\bf \nabla}^2 {\bf u}\right) ^2 dV \leq {\mathcal G}  \sqrt{ \int {\mathbf \omega}^2 dV} \, \sqrt{ \int \left( {\bf \nabla}\times{\bf f}\right) ^2 dV}. \label{ineq}
\end{equation}
Applying Young's inequality for any positive real number $a$ provides the following expression:
\begin{equation}
H\!a  \int {\mathbf \omega}^2 dV + \int \left( {\bf \nabla}^2 {\bf u}\right) ^2 dV \leq \frac{{\mathcal G}}{2\, a} \int {\mathbf \omega}^2 dV \, + \, \frac{{\mathcal G}\, a}{2} \int \left( {\bf \nabla}\times{\bf f}\right) ^2 dV. \label{ineq2}
\end{equation}
The optimal choice $a=0.5\, {\mathcal G} \, H\!a^{-1}$ enables us to take advantage of the enstrophy term on the left hand side of (\ref{ineq2}), and one can finally obtain the following result:
\begin{equation}
\overline{C} \ =\ \sqrt{ \int \left( {\bf \nabla}^2 {\bf u}\right) ^2 dV } \leq \ \frac{{\mathcal G}}{2 \sqrt{H\!a}} \sqrt{ \int { \left( {\bf \nabla}\times{\bf f}\right) ^2 dV}}. \label{ineq4}
\end{equation}
If one restrict ones attention to large-scale forcing, then the dimensionless ${\bf \nabla}\times{\bf f}$ is of order unity, as the dimensionless forcing ${\bf f}$ and large scales are of order unity. Hence inequality (\ref{ineq4}) becomes:
\begin{equation}
\overline{C} \ =\ \sqrt{ \int \left( {\bf \nabla}^2 {\bf u}\right) ^2 dV } \leq \ \frac{{\mathcal G}}{2 \sqrt{H\!a}} . \label{ineq5}
\end{equation}

}





\end{document}